\documentclass{article}

\usepackage{PRIMEarxiv}

\usepackage[utf8]{inputenc} % allow utf-8 input
\usepackage[T1]{fontenc}    % use 8-bit T1 fonts
\usepackage{hyperref}       % hyperlinks
\usepackage{url}            % simple URL typesetting
\usepackage{booktabs}       % professional-quality tables
\usepackage{amsfonts}       % blackboard math symbols
\usepackage{nicefrac}       % compact symbols for 1/2, etc.
\usepackage{microtype}      % microtypography
\usepackage{lipsum}
\usepackage{fancyhdr}       % header
\usepackage{graphicx}       % graphics
\graphicspath{{media/}}     % organize your images and other figures under media/ folder
\usepackage{authblk}
\usepackage[table,dvipsnames]{xcolor}
\usepackage{multirow}%
\usepackage{amsmath,amssymb}%
\usepackage{amsthm}%
\usepackage{mathrsfs}%
\usepackage[title]{appendix}%
\usepackage{xcolor}%
\usepackage{textcomp}%
\usepackage{manyfoot}%
\usepackage{algorithm}%
\usepackage{algorithmicx}%
\usepackage{algpseudocode}%
\usepackage{braket}
\usepackage{listings}%
\usepackage{subcaption} % for subfigures
\usepackage{mathtools}

\title{Quantum-Assisted Design of Space-Terrestrial Integrated Networks}

\author[1,2]{Chiara Vercellino}
\author[1,2]{Giacomo Vitali}
\author[1]{Paolo Viviani}
\author[1]{Alberto Scionti}
\author[1]{Olivier Terzo}
\author[2]{Bartolomeo Montrucchio}
\author[3]{Pascal Jahan Elahi}
\author[3]{Ugo Varetto}

\affil[1]{Fondazione LINKS, Torino, IT \\ \texttt{firstname.lastname@linksfoundation.com}}
\affil[2]{DAUIN, Politecnico di Torino, Torino, IT \\ \texttt{firstname.lastname@polito.it}}
\affil[3]{Pawsey Supercomputing Research Centre, Perth, AUS}

\begin{document}
\maketitle

\begin{abstract}
Achieving ubiquitous global connectivity requires integrating satellite and terrestrial networks, particularly to serve remote and underserved regions. In this work, we investigate the design and optimization of Space-Terrestrial Integrated Networks (STINs) using a hybrid quantum-classical approach. We formalize three key combinatorial optimization problems: the Satellite Selection Problem (SSP), the Gateway Selection Problem (GSP), and the Spectrum Assignment Problem (SAP), each capturing critical aspects of network deployment and operation. Leveraging neutral-atom quantum processors, we map the SSP onto a Maximum Weight Independent Set problem, embedding it onto the \textit{Aquila} platform and solving it via the Quantum Adiabatic Algorithm (QAA). Postprocessing ensures feasible solutions that guide downstream GSP and SAP optimization. Benchmarking across 165 realistic remote regions shows that QAA solutions closely match classical exact solvers and outperform greedy heuristics, while subsequent GSP and SAP outcomes remain largely robust to differences in initial satellite selection. These results demonstrate that quantum optimization achieves performance broadly comparable to classical approaches for end-to-end STIN design, with rare instances where it can even surpass state-of-the-art solvers. This suggests that, while not yet consistently superior, quantum methods may offer competitive advantages for larger or more complex instances of the underlying combinatorial subproblems.
\end{abstract}

% keywords can be removed

\section{Introduction}\label{sec:intro}

With global communication demands continuing to grow, extending connectivity to rural and remote areas has become a key objective for future networks. The evolution of wireless communication has consistently pursued more efficient, reliable, and ubiquitous services. The upcoming sixth-generation (6G) network is expected to revolutionize the communication landscape by enabling seamless integration of heterogeneous services and ensuring continuous coverage worldwide. In this context, the integration of satellite and terrestrial networks has emerged as a promising solution to deliver global broadband access, attracting significant attention from both academia and industry~\cite{zhu2021integrated}.  

From the first generation (1G) to the fifth generation (5G), terrestrial wireless communication has achieved major advances in speed, latency, and overall quality of service~\cite{rost2016mobile}. High-speed broadband with low latency is now widely accessible within the coverage of terrestrial Base Stations (BSs). However, due to economic and geographical constraints, terrestrial infrastructure deployment has focused primarily on developed and densely populated areas~\cite{kuang2017radio}. As a result, vast rural and remote regions remain underserved, limiting equitable access to modern communication services. Despite its improvements, 5G still falls short of providing universal connectivity, with large portions of the global population remaining unconnected~\cite{yaacoub2019key}. Natural barriers such as oceans, mountains, and sparsely populated territories further hinder expansion. To address these challenges, 6G must shift its focus from merely improving performance metrics to achieving ubiquitous global coverage.  

Satellite networks offer a viable solution to these limitations. Their extensive coverage enables connectivity across wide geographic areas at lower cost than large-scale terrestrial deployments~\cite{su2019broadband}. Unlike ground-based networks, satellite systems provide top-down coverage largely unaffected by geography, making them particularly suitable for rural, maritime, and airborne applications. Historically, large-scale satellite communication faced challenges such as high deployment costs and limited transmission capacity. However, recent advances in satellite technology and the rise of mega-constellations (e.g., SpaceX Starlink, OneWeb, Telesat) have revitalized interest in satellite broadband services~\cite{del2019technical}. Nevertheless, satellites alone cannot fully replace terrestrial infrastructure in dense urban regions, where ground-based networks remain more cost-effective. Instead, a hybrid architecture that integrates satellite and terrestrial systems provides a complementary approach that leverages the strengths of both.  

A Space-Terrestrial integrated network (STIN) represents the next evolutionary step in wireless communication. By combining the advantages of both architectures, STIN aims to deliver seamless, high-quality connectivity on a global scale. Compared to conventional terrestrial systems, STIN offers several advantages: (i) extended rural and remote coverage, ensuring reliable connectivity in sparsely populated, maritime, and airborne environments; (ii) enhanced reliability through joint use of satellite and terrestrial components, enabling service continuity during infrastructure failures; (iii) efficient broadcast and multicast services, optimizing content delivery and reducing redundancy~\cite{montalban2018multimedia}; and (iv) resilience to natural disasters, where satellite connectivity can sustain communication when terrestrial BS and fiber infrastructure are disrupted.  

As the world moves toward 6G, STIN will play a crucial role in bridging the digital divide and enabling a truly connected world. However, integrating these heterogeneous technologies into a cohesive, high-performance system introduces substantial challenges. One key difficulty lies in optimizing the roles and coordination of different components, including ground stations, gateways, and satellites, particularly to enhance connectivity in underserved regions.  

The QRISTIN project (Quantum Routines In Space-Terrestrial Integrated Networks) directly targets some of these challenges. Its objective is to formulate and solve fundamental optimization problems associated with STIN design \& deployment. As problem instances grow in size and structural complexity, the computational effort required by conventional optimization methods generally increases. While small-scale instances may be solved efficiently, larger or more densely connected problems can demand substantially more resources.

QRISTIN explores the potential of quantum computing as an enabler for next-generation communication networks. By leveraging hybrid quantum-classical algorithms, QRISTIN investigates how quantum resources can either accelerate classical workflows or serve as novel solution paradigms for complex optimization tasks.  

The remainder of this paper is organized as follows. Section~\ref{sec:prob_def} introduces the optimization problems relevant to STIN identified within QRISTIN. Section~\ref{sec:method} presents the optimization methodology, focusing on hybrid quantum-classical algorithms. Section~\ref{sec:results} reports the results obtained using the \textit{Aquila} quantum processing unit (QPU), benchmarked against fully classical approaches (exact and heuristic). Finally, Section~\ref{sec:conclusion} concludes the paper by summarizing the key findings and highlighting potential future research directions.  

\section{Problem Definitions}\label{sec:prob_def}

This section introduces the optimization problems arising in the design and operation of STINs. These problems are computationally hard and exhibit graph-based structures, which makes them appealing testbeds for quantum optimization techniques, especially those exploiting neutral atoms hardware. In the context of the QRISTIN project, we formalize three such problems:

\begin{enumerate}
    \item \emph{Satellite Selection Problem (SSP)}, where the objective is to determine which satellites should be activated to provide optimal coverage in underserved regions;
    \item \emph{Gateway Selection Problem (GSP)}, which concerns the efficient association of satellites with terrestrial gateways to ensure load balancing and reliable inter-layer connectivity;
    \item \emph{Spectrum Assignment Problem (SAP)}, which addresses the allocation of spectrum resources across the integrated network to minimize interference.
\end{enumerate}

Before presenting the formulations, we state two common modeling assumptions: \textit{i)} for the satellite component, we focus on low Earth orbit (LEO) constellations~\cite{ma2024leo}, motivated by their low latency, publicly available orbital datasets enabling realistic modeling, and their central role in current and planned constellations; \textit{ii)} we consider a multi-service system configuration in which user devices are dual-mode, capable of connecting to both terrestrial and satellite layers. This allows seamless switching and combined use of resources across the two domains.

\subsection{Satellite Selection Problem (SSP)}

The \emph{Satellite Selection Problem} addresses the task of extending terrestrial coverage in remote or underserved areas by selecting an appropriate subset of LEO satellites. Given a service area $\Omega$, let $\Omega' \subset \Omega$ denote the portion not covered by terrestrial base stations. The objective is to maximize coverage of $\Omega'$ while avoiding redundant overlap between satellites.  

Formally, let:
\begin{itemize}
  \item $\Omega$ be the total service area;
  \item $\Omega' \subset \Omega$ the uncovered portion;
  \item $S = \{s_1, \dots, s_n\}$ the set of candidate LEO satellites;
  \item $w_i$ the weight of satellite $s_i$, proportional to the area it covers in $\Omega'$ over a defined time window;
  \item $E \subseteq S \times S$ the set of edges representing overlapping satellite coverage exceeding a threshold $\epsilon=0.90$;
  \item $x_i \in \{0,1\}$ a binary variable indicating whether $s_i$ is selected.
\end{itemize}

The sets $S$, $E$, and $\{w_i\}$ define a weighted undirected graph on which the problem reduces to a Maximum Weight Independent Set (MWIS). The goal is to identify an independent subset of satellites (no two with excessive overlap) that maximizes coverage. This NP-hard problem can be expressed as the following integer linear program (ILP):

\begin{align}
\text{maximize} \quad &\sum_{i \in S} w_i x_i, \\
\text{subject to} \quad &
\begin{cases} 
&x_i + x_j \leq 1 \quad \forall (s_i,s_j) \in E, \\
&x_i \in \{0,1\} \quad \forall s_i \in S,
\end{cases}\quad.
\end{align}

Variants of the SSP extend the formulation to cases where satellites substitute, rather than complement, terrestrial BSs. Relevant applications include \emph{energy-efficient infrastructure management}, where BS can be deactivated if satellite coverage suffices, and \emph{elastic capacity extension}, where satellites temporarily augment coverage in areas with sudden demand surges (e.g., mass events).  

\subsection{Gateway Selection Problem (GSP)}

The \emph{Gateway Selection Problem} concerns the association of satellites with terrestrial gateways to ensure inter-layer connectivity. Each satellite must be connected to one gateway, while the load should be balanced across gateways. Extensions of the formulation may include satellite gateways or cost functions to capture heterogeneity among links.  

Let:
\begin{itemize}
  \item $G = \{g_1, \dots, g_m\}$ be the set of terrestrial gateways;
  \item $L \subseteq S \times G$ the set of feasible satellite--gateway links;
  \item $y_{ij} \in \{0,1\}$ a binary variable equal to one if $s_i$ is assigned to $g_j$;
  \item $M$ the maximum number of satellites assigned to any single gateway;
  \item $c_{ij}$ optional link-specific costs, such as latency, propagation, or cost considerations, yielding weighted variants of the GSP.
\end{itemize}

The optimization objective is to minimize $M$, thereby promoting balanced gateway loads. The corresponding ILP is:

\begin{align}
\text{minimize} \quad & M, \\
\text{subject to} \quad & 
\begin{cases} \label{eq:gsp}
    & \sum_{j: (s_i,g_j) \in L} y_{ij} = 1 \quad \forall s_i \in S, \\
    & \sum_{i: (s_i,g_j) \in L} y_{ij} \leq M \quad \forall g_j \in G, \\
    & y_{ij} \in \{0,1\} \quad \forall (s_i,g_j) \in L, \\
    & M \in \{0,1,\ldots, n\}, 
\end{cases}\quad.
\end{align}

\subsection{Spectrum Assignment Problem (SAP)}

The \emph{Spectrum Assignment Problem} arises once the network connectivity topology is determined by the SSP and GSP solutions. The goal is to allocate spectrum resources to communication paths in a manner that minimizes interference, a problem naturally modeled as graph coloring.  

Formally, let:
\begin{itemize}
  \item $P = \{p_1, \dots, p_h\}$ be the set of logical transmission paths;
  \item $L = \{l_1, \dots, l_k\}$ the set of physical communication links;
  \item $E \subseteq P \times P$ the set of edges between paths sharing at least one physical link;
  \item $K$ the set of available frequency bands;
  \item $x_{pq} \in \{0,1\}$ a binary variable equal to one if path $p$ is assigned frequency $q$;
  \item $c_{pq}$ denotes the cost of assigning frequency $q$ to path $p$, which may be uniform or scenario-dependent (e.g., regulatory or propagation constraints).  
\end{itemize}

Then, we can formulate the corresponding ILP as follow:

\begin{align}
\text{minimize} \quad & \sum_{p \in P} \sum_{q \in K} c_{pq} \cdot x_{pq}, \\
\text{subject to} \quad & 
\begin{cases}
    &\sum_{q \in K} x_{pq} = 1 \quad \forall p \in P, \\
    & x_{pq} + x_{rq} \leq 1 \quad \forall (p, r) \in E, \forall q \in K, \\
    & x_{pq} \in \{0,1\} \quad \forall p \in P, \forall q \in K,
\end{cases}\quad.
\end{align}

Although spectrum assignment is related to routing, we treat routing as a separate stage. For isolated requests, shortest-path algorithms suffice, while simultaneous multi-commodity flows require more complex formulations. Classical algorithms (e.g., Dijkstra’s~\cite{javaid2013understanding}, Bellman-Ford~\cite{magzhan2013review}, Edmonds-Karp~\cite{edmonds1972theoretical}) already provide efficient solutions for routing; hence, the SAP represents a more natural candidate for quantum optimization.

\section{Methodology}\label{sec:method}

The optimization problems introduced in Section~\ref{sec:prob_def}, while different in scope, share a unifying structure: each can be expressed as a graph-based combinatorial problem. In these representations, vertices correspond to network elements such as satellites, gateways, or communication paths, while edges encode constraints including overlaps, connectivity, or interference. This abstraction not only provides a common ground for classical optimization but also creates a direct link to quantum solution strategies.

Among the various quantum computing platforms, neutral-atom quantum processors have recently emerged as promising candidates for solving graph-based problems at scale~\cite{serret2020solving}. In such systems, individual atoms are trapped in optical tweezers and manipulated using laser pulses. When excited into Rydberg states, atoms strongly interact with nearby atoms, giving rise to the so-called \emph{Rydberg blockade} effect. This effect prevents two neighboring atoms from being simultaneously excited, thereby naturally enforcing independent set constraints~\cite{adams2019rydberg}. Additional information on the dynamics of the quantum system can be found in Appendix \ref{app:hamiltonian}.

This physical mechanism establishes a direct correspondence between the device and the Maximum (Weight) Independent Set (M(W)IS) problem. Low-energy configurations of the atomic system map to feasible independent sets, and weighted formulations allow coverage or capacity considerations to be integrated. This property is central to our study: the SSP maps directly to an MWIS instance, and its outcomes provide the structural basis to address the GSP and SAP. Thus, MWIS serves as a unifying computational primitive for the STIN optimization tasks identified in the QRISTIN project.

To exploit this correspondence in practice, the abstract optimization problem must first be \emph{embedded} into the quantum hardware. This requires arranging atoms in a two-dimensional array such that the interaction pattern reflects the adjacency structure of the target graph. Once an embedding is defined, algorithms such as the Quantum Adiabatic Algorithm (QAA)~\cite{albash2018adiabatic} can be used to guide the system towards low-energy states representing candidate solutions.

As is common with near-term devices, the quantum output is approximate, and affected by noise and hardware limitations. To mitigate these effects, classical postprocessing is applied to refine the raw quantum results, improve feasibility, and enhance solution quality. The overall workflow is therefore hybrid in nature, combining quantum exploration of the solution space with classical optimization for fine-tuning, thereby balancing the strengths of both approaches.

\subsection{Embedding Optimization Problems on the QPU}\label{sec:embedding}

To execute quantum optimization algorithms on the chosen neutral-atom platform \textit{Aquila}~\cite{wurtz2023aquila}, the first requirement is to embed the target problem instance into the physical qubit register. Formally, we consider an input graph $\mathcal{G}(\mathcal{V}, \mathcal{E})$, where vertices $\mathcal{V}$ represent network elements and edges $\mathcal{E}$ encode pairwise constraints. The embedding task consists of mapping $\mathcal{G}$ to a unit disk graph (UDG)~\cite{clark1990unit}, where each vertex $i \in \mathcal{V}$ is associated with a physical position $\vec{r}_i \in \mathbb{R}^2$ corresponding to an atom in the two-dimensional register. This mapping must respect the geometric limitations of the hardware, summarized in Table~\ref{tab:hardware-constraints}.

\begin{table*}[h]
\centering
\caption{Geometric constraints of the \textit{Aquila} QPU register.}
\label{tab:hardware-constraints}
\begin{tabular}{|l|l|}
\hline
\textbf{Constraint} & \textbf{Requirement} \\
\hline
Minimum vertical row spacing ($d_{\text{row}}$) & $\geq 2 \,\mu\text{m}$ \\
Minimum inter-atom distance ($D_{\min}$) & $\geq 4 \,\mu\text{m}$ \\
Register dimensions ($L_x\times L_y$) & $76 \,\mu\text{m} \times 128 \,\mu\text{m}$ \\
\hline
\end{tabular}
\end{table*}

To address the embedding challenge, we use the Distance Encoder Network (DEN) model~\cite{vercellino2023neural,vercellino2024harnessing}, adapted to the rectangular geometry of \textit{Aquila}. The model learns a nonlinear transformation from an initial set of coordinates $\{\vec{r}_i^{\,0}\}$ to a feasible configuration $\{\vec{r}_i\}_{i\in \mathcal{V}}$ that satisfies hardware constraints while preserving the adjacency structure of the graph.

The DEN is based on a modified autoencoder, where a hidden \emph{coordinates layer} encodes feasible atomic positions. Outputs are evaluated in terms of pairwise and row distances, and optimized through a custom embedding loss function (ELF). This loss enforces the UDG constraints while promoting a large \emph{adjacency gap}, i.e., the separation between the closest non-adjacent pair ($D$) and the most distant adjacent pair ($d$). In a proper UDG setting, we have $d<D$. Extensive details about the embedding methodology can be found in Appendix \ref{app:embedding}.

The workflow operates in three stages:
\begin{enumerate}
    \item \textbf{Preprocessing:} generate an initial embedding $\vec{r}_i^{\,0}$ for all $i \in \mathcal{V}$;  
    \item \textbf{Learning:} map $\{\vec{r}_i^{\,0}\}_{i\in \mathcal{V}}$ into a feasible embedding $\vec{r}_i$ using the DEN model;  
    \item \textbf{Postprocessing:} refine $\{\vec{r}_i\}_{i\in \mathcal{V}}$ with the L-BFGS-B optimizer within a restricted solution space.
\end{enumerate}

\textbf{Initialization strategy:} Since the DEN model requires initial coordinates, we employ the Fruchterman–Reingold force-directed layout algorithm~\cite{SciPyProceedings_11, fruchterman1991graph}. This algorithm balances attractive forces between adjacent vertices and repulsive forces between all vertices, producing layouts that approximately satisfy spacing constraints. We set the force parameter to $k = 7\,\mu$m, consistent with the device’s feasible distance range for $d$ ($4$–$10\,\mu$m). While this initialization does not guarantee valid embeddings, it accelerates convergence during training.

\textbf{Neural optimization and loss formulation:} The DEN improves the initial positions using an autoencoder architecture combined with a distance-computation block. This block ensures that outputs are interpreted as Cartesian coordinates, with squared pairwise and row distances serving as inputs to the ELF. The ELF integrates four objectives:  

\begin{itemize}
    \item minimum-distance enforcement $D_{\min}$,  
    \item distance control between adjacent vertices ($d$) and non-adjacent vertices ($D$),  
    \item row-spacing enforcement $d_{\text{row}}$, and  
    \item adjacency-gap maximization ($D - d$), to pursue a UDG representation of $\mathcal{G}$.  
\end{itemize}

The final loss is a weighted sum of these objectives, with relaxed weighting for larger graphs to improve convergence. Training is instance-specific, runs for up to 5000 epochs, and employs dropout regularization together with the AdamW optimizer.

\textbf{Refined adjustment:} After obtaining the embedding from the DEN model, vertex positions are further improved through a constrained continuous optimization step using L-BFGS-B~\cite{zhu1997algorithm}. Coordinates are first translated so that the layout's lower-left corner aligns with $(0,0)$ and rounded to a fixed grid. Safe bounding boxes are computed for each vertex to enforce minimum Euclidean distance ($d_{\min}=4\,\mu$m), row spacing ($d_{\text{row}}=2\,\mu$m), and register area limits. The optimization refines vertex positions within these feasible regions using a margin-based loss that penalizes adjacent vertices placed too far apart and non-adjacent vertices placed too close. Validation is performed by comparing $d$ with $D$; if $d < D$, the graph admits a valid unit-disk embedding. The resulting layouts are feasible on \textit{Aquila} and ready for deployment.

\subsection{Quantum Optimization with Aquila}

Once refined embeddings are obtained, vertex positions are directly mapped to qubit locations on the \textit{Aquila} QPU. Each vertex corresponds to a neutral atom, whose $x$-$y$ position in the tweezer array is given by the embedding $\{\vec{r}_i\}_{i\in \mathcal{V}}$. This spatial arrangement allows adjacency constraints to be enforced naturally through the Rydberg blockade, providing a physical realization of graph-based optimization problems.

\subsubsection{Quantum Adiabatic Algorithm for MWIS}

The MWIS problem, central to the SSP, is solved using the Quantum Adiabatic Algorithm (QAA)~\cite{serret2020solving}. Vertex weights are encoded via site-specific \emph{detunings} of each qubit, while adjacency constraints are enforced by the Rydberg blockade between nearby atoms~\cite{bombieri2025quantum}. The blockade radius $r_b$ determines the interaction range and is chosen based on the embedding distances: $r_b = \sqrt{d \cdot D}$~\cite{matwiejew2025continuous}. This choice ensures a valid physical encoding for proper unit-disk instances while maintaining feasible control parameters for other embeddings.

The \textit{Aquila} hardware provides two detuning channels: a global detuning and a local, vertex-specific detuning. The global channel defines the baseline sweep of the quantum evolution, while the local channel encodes vertex weights into energy shifts. Weights are rescaled so that heavier vertices are energetically favored, biasing the dynamics toward high-weight independent sets. The Rabi frequency and global detuning are varied piecewise-linearly in time to implement the adiabatic schedule. For small graphs ($|\mathcal{V}|\leq15$), the QAA can be simulated locally with classical computational resources; larger graphs are run on \textit{Aquila}’s QPU.

\subsubsection{Postprocessing and Solution Refinement}\label{sec:postproc}

The QAA output is probabilistic, returning bitstrings that encode candidate independent sets. Each bitstring is interpreted according to \textit{Aquila}’s measurement convention: empty traps correspond to selected vertices, and occupied traps correspond to excluded vertices. Bitstrings corresponding to non-independent set of vertices in $\mathcal{G}$ are discarded, and a greedy augmentation adds unselected vertices in descending weight order whenever possible. This procedure yields feasible solutions that approximate maximum-weight independent sets, compensating for both hardware noise and imperfect embeddings. By combining quantum sampling with classical refinement, this hybrid approach produces reliable, high-quality solutions to the SSP.

\section{Results}\label{sec:results}

To benchmark the quantum optimization method (QAA) against classical approaches, we constructed a dataset of STIN instances covering 165 remote regions, defined as areas with limited terrestrial coverage ($\leq 60\%$ of the total territory). Terrestrial BS and gateway locations were extracted from OpenStreetMap\footnote{\url{https://wiki.openstreetmap.org/wiki/Map_features\#Telecom}}, while satellite data were obtained from Celestrak\footnote{\url{https://celestrak.org/NORAD/elements/}}, selecting the ONEWEB constellation.  

For each satellite, we simulated its orbit over a complete orbital period and identified the time intervals during which it covered the target regions. From this information, we generated instances for each STIN subproblem:  
\begin{itemize}
    \item \textbf{SSP instances:} graphs where vertices represent satellites, vertex weights correspond to the percentage of the covered area (excluding portions already served by terrestrial BSs) averaged over the time window, and edges connect satellites whose coverage overlaps by more than 90\% over the same region at the same time.  
    \item \textbf{GSP instances:} bipartite graphs with vertices representing satellites or terrestrial gateways, and edges indicating feasible communication links between them.  
    \item \textbf{SAP instances:} graphs defined by transmission paths connecting satellites and reachable BSs via gateways. Each path is a vertex, and edges indicate shared links between paths.
\end{itemize}

For the SSP, we compared three solution methods:
\begin{enumerate}
\item The quantum optimization approach (QAA) described in Section~\ref{sec:method}. To balance resource usage with exploration of the solution space, we fixed the number of quantum measurements at 300.
\item An exact ILP solver using GLPK~\cite{glpk}, implemented via the \texttt{Pyomo} library~\cite{hart2017pyomo}. Both quantum and classical approaches were evaluated  in terms of solution quality under a fixed walltime constraint. The ILP solver was limited to 60 seconds of execution, matching the time required for 300 quantum measurements sampled at 5 Hz.
\item A greedy heuristic, which iteratively selects the satellite with the largest weight until no additional satellite can be added without violating the independence constraint. This heuristic can be regarded as the \textit{no-quantum limit}~\cite{wurtz2024solving}, as it effectively corresponds to an all-null bitstring to which we applied the postprocessing described in Section~\ref{sec:postproc}. Including this heuristic as a benchmark provides insights into the effective quantum utility~\cite{wurtz2021classically}.
\end{enumerate}

% \begin{figure}[ht!]
%     \centering
%     \includegraphics[width=1\linewidth]{graphics/SSP_boxplot_grouped_by_vertices.pdf}
%     \caption{SSP objective values grouped by problem size $|\mathcal{V}|$, comparing the greedy heuristic, exact ILP solver, and QAA method. \PJE{What is the distribution over? Purely $|\mathcal{V}|$ or are there several instances per given $|\mathcal{V}|$. Why are the points at large problem instance just lines?} } \CV{yes, the intervals are bins for the number of vertices, but in the last bins I just have one sample so it is just one point. If you think this is not relevant I can just remove the figure and the reference to it, so it would also be easier to stay in the 12 page limit}
%     \label{fig:ssp_plot}
% \end{figure}

\begin{figure}[ht!]
    \centering
    \includegraphics[width=1\linewidth]{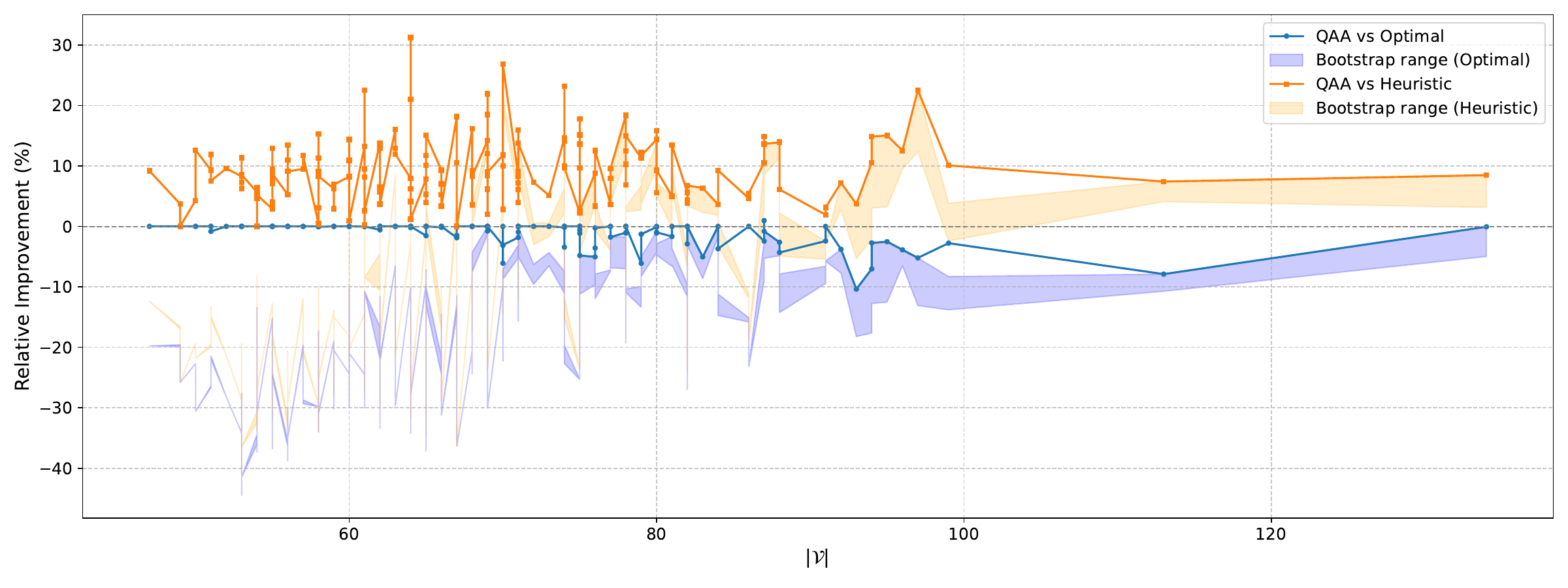}
    \caption{Relative improvement of QAA compared to the greedy heuristic (orange) and exact solver (blue) for each graph instance.}
    \label{fig:ssp_rel_imp}
\end{figure}

% Figure~\ref{fig:ssp_plot} shows the distribution of objective values across the dataset for each method. The QAA closely matches the exact solver, while the greedy heuristic underperforms both.

Figure~\ref{fig:ssp_rel_imp} presents the relative performance of QAA versus the heuristic and the exact solver. We observe that QAA consistently outperforms the heuristic, with relative improvement increasing with problem size $|\mathcal{V}|$, although very large graphs ($|\mathcal{V}|>100$) are few. On average, QAA achieves a $9.2\%$ improvement over the heuristic, and a $-0.7\%$ difference compared to the exact solver. We observed one instance in which the QAA outperformed the GLPK exact solver. For a graph with 96 vertices, the quantum approach achieved an objective value of 16.50, compared to 16.22 for the classical solution. This result is interesting, as the embedding of the corresponding graph is not unit-disk. This highlights the potential for quantum methods to occasionally surpass state-of-the-art classical solvers, even within constrained time limits, and under non-perfect problem representations.
To provide insights into the effect of quantum measurements on the final objective values, in Fig.~\ref{fig:ssp_rel_imp} we also report bootstrap regions, obtained by extracting 100 bitstrings out of 300 and performing 20 bootstraps on the original distributions to compute the objective values. The bootstrapped solutions consistently underperform, often yielding worse results even compared to the heuristic approach, particularly for instances with a lower number of vertices $|\mathcal{V}|$. For larger graphs, the performance gap is reduced, suggesting that while 300 measurements are sufficient to adequately explore the solution space in smaller instances, a larger number of measurements is likely needed for larger problem sizes.

To analyze the impact of the solution refinement procedure, we computed the \textit{Hamming distance}, normalized by the number of vertices, between the original and post-processed bitstrings. This provides a metric for evaluating how many vertices are added, on average, during post-processing. Overall, the refinement step increases the number of vertices by approximately $10\%$ on average, with a standard deviation of about $0.05$. For two instances, the percentage was $0\%$, meaning no modifications were required, while in the worst case this value reached $26\%$.

\begin{figure}[ht!]
    \centering
    \includegraphics[width=1\linewidth]{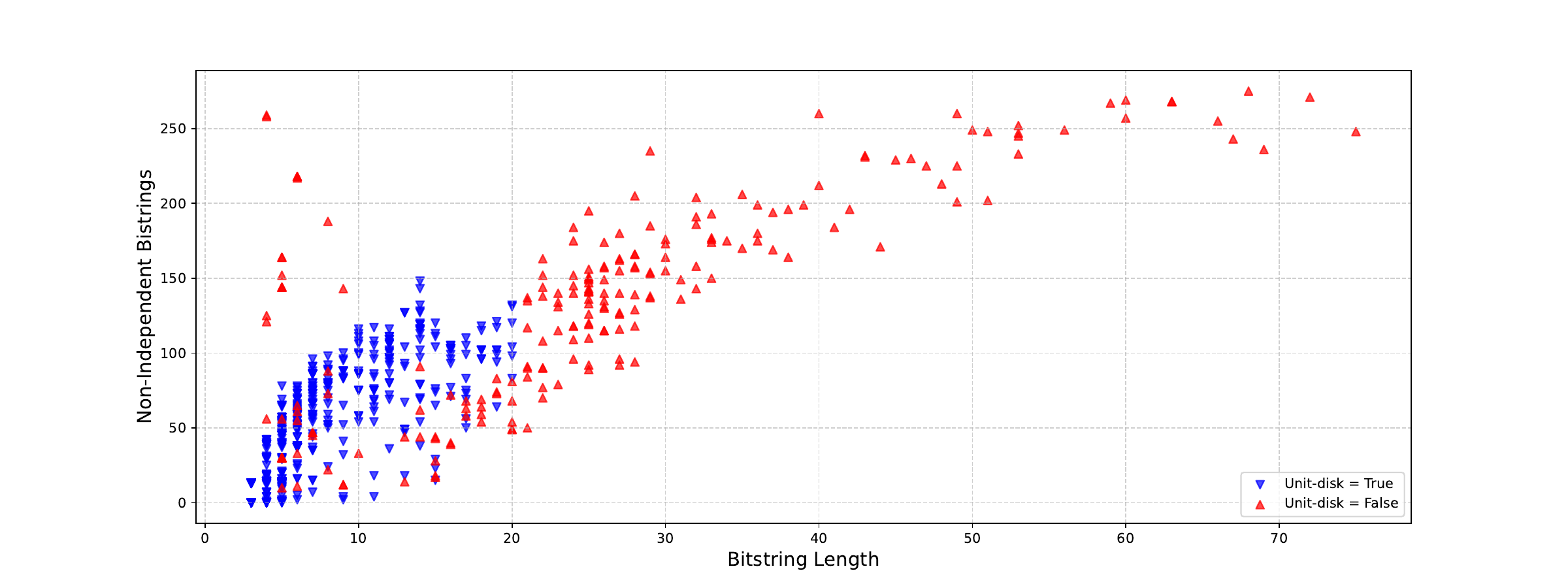}
    \caption{Scatter plot showing the number of non-independent bitstrings as a function of the number of vertices.}
    \label{fig:nonindep}
\end{figure}

Another important metric is the number of \emph{useful bitstrings}, i.e., those that do not correspond to vertex subsets that are not independent (which we call \textit{non-independent bitstrings}). Figure~\ref{fig:nonindep} shows the occurrences of these non-independent bitstrings across all graph samples in our dataset, considering a fixed total of $300$ measurements for each graph. We observe that the number of non-independent bitstrings tends to increase with the graph size, and is also influenced by whether or not the graph can be represented as a UDG in the QPU embedding. Larger graphs are generally harder to embed as UDGs, which in turn affects the representation of adjacency patterns among vertices and ultimately impacts the overall solution quality.

\begin{figure}[ht!]
    \centering
    
    % First subfigure
    \begin{subfigure}{0.48\textwidth}
        \centering
        \includegraphics[width=\linewidth]{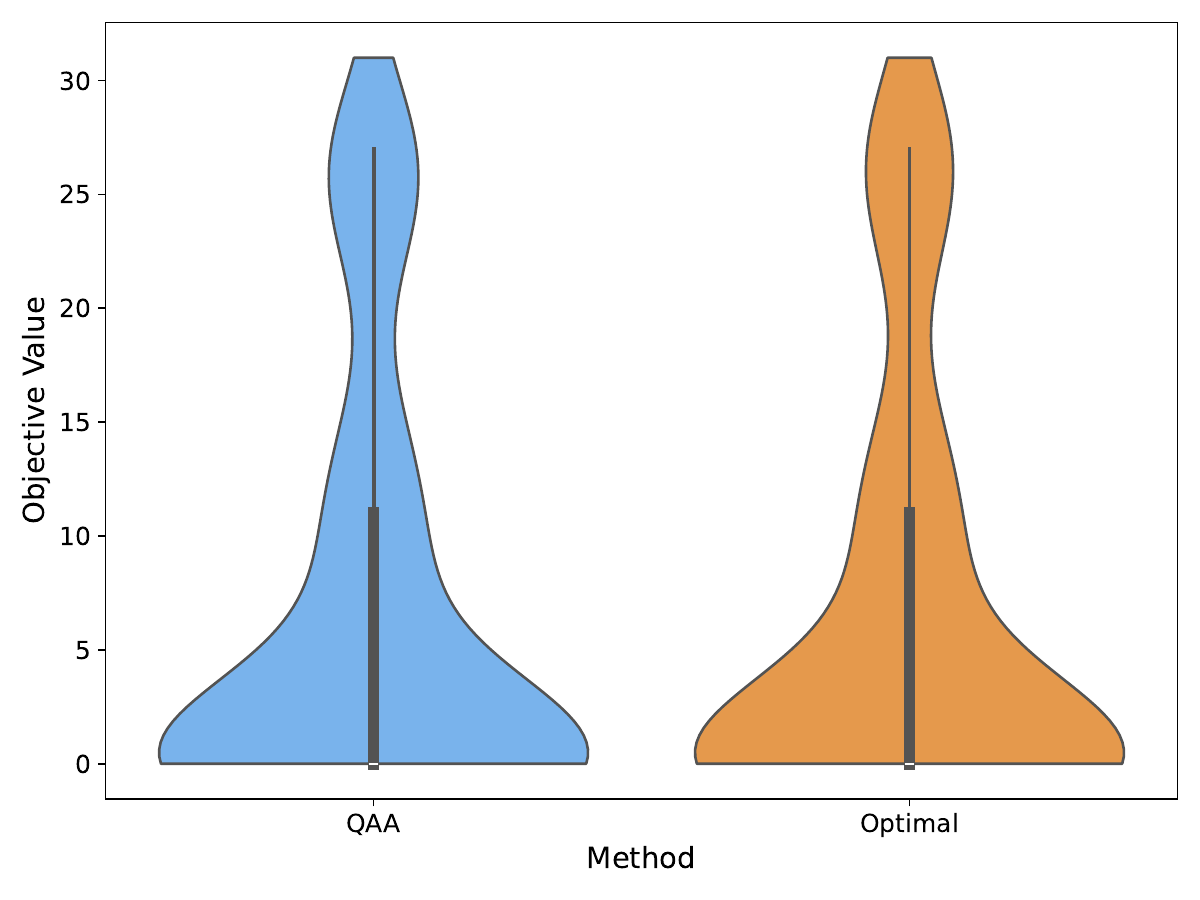}
        \caption{Distribution of GSP objective values using satellite sets obtained from SSP by either QAA or the exact solver.}
        \label{fig:gsp_plot}
    \end{subfigure}
    \hfill
    % Second subfigure
    \begin{subfigure}{0.48\textwidth}
        \centering
        \includegraphics[width=\linewidth]{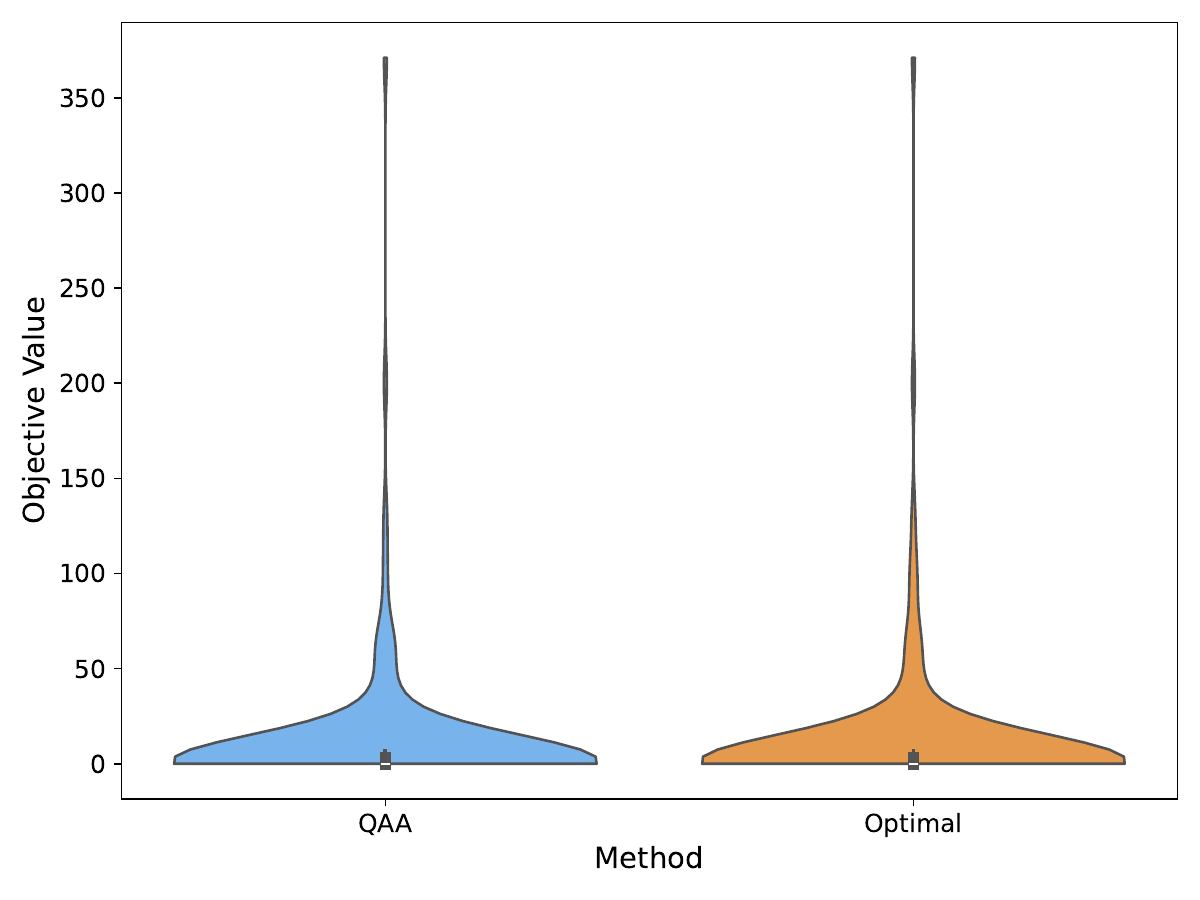}
        \caption{Distribution of SAP objective values based on network topologies derived from preceding SSP and GSP solutions.}
        \label{fig:violinplot}
    \end{subfigure}
    
    \caption{Comparison of GSP and SAP objective distributions when varying the SSP input satellites obtained from either the exact solver or the quantum method (QAA).}
    \label{fig:comparison}
\end{figure}

Although the SSP solutions obtained by QAA and the exact solver differed in 94 out of 165 instances, 51 affecting the objective function and 43 yielding alternative satellite combinations with the same objective, the resulting GSP and SAP solutions are largely consistent. Figure~\ref{fig:comparison} illustrates the distribution of objective values for both problems, showing that variations in the initial satellite selection have only a limited impact on subsequent optimization stages. For the GSP case, the Shannon-Jensen divergence for the GSP case is 0.00013, with only 9 out of 165 instances exhibiting different objectives; in all these cases, the input derived from the preceding SSP led to a lower (i.e., more balanced) distribution when solved with QAA rather than with the exact solver. The distributions divergence is larger in the SAP case (0.00218), where only 6 instances showed differences in objective values, but the gap was more pronounced, with solutions coming from a fully-classical pipeline requiring up to 35 additional frequency bands compared to those derived from QAA.

Overall, these results indicate that, despite differences in the SSP solutions between QAA and the exact solver, the subsequent GSP and SAP outcomes are largely robust to the choice of input satellites. This suggests that the quantum approach can achieve practically equivalent performance for end-to-end STIN optimization, and at larger scales may offer a competitive alternative to classical methods for the underlying combinatorial subproblems.

\section{Conclusion}\label{sec:conclusion}

In this work, we explored the application of hybrid quantum-classical optimization to Space-Terrestrial Integrated Networks (STINs), focusing on the challenges of providing reliable connectivity to remote and underserved regions. We formalized three combinatorial problems—Satellite Selection (SSP), Gateway Selection (GSP), and Spectrum Assignment (SAP)—which collectively define the network optimization pipeline. By mapping the SSP to a Maximum Weight Independent Set problem, embedding it onto \textit{Aquila} neutral-atom quantum processor, and solving it via the Quantum Adiabatic Algorithm (QAA), we demonstrated that quantum approaches can produce solutions closely matching classical exact solvers and consistently outperforming greedy heuristics.  

Our results show that variations in SSP solutions have limited impact on subsequent GSP and SAP outcomes, indicating that end-to-end STIN optimization is largely robust to the choice of initial satellite selection. This suggests that quantum-assisted solutions can be effectively integrated into the overall network design workflow without significant downstream performance degradation.  

Although current neutral-atom quantum hardware imposes limits on the size of solvable instances, these findings indicate that, at larger scales, quantum optimization may serve as a viable alternative or complement to classical methods for computationally intensive network design problems. Future research directions include extending the problem formulations with additional operational constraints, testing the quantum algorithms on different hardware architectures, and exploring alternative quantum optimization strategies, such as the Quantum Approximate Optimization Algorithm (QAOA)~\cite{choi2019tutorial} with pulse shaping, to further improve performance and scalability.

\section*{Acknowledgments}
This work was supported by the NGI Sargasso initiative under the Horizon Europe programme (Grant Agreement No. 101092887). This work was also supported by resources provided by the Pawsey Supercomputing Research Centre with funding from the Australian Government and the Government of Western Australia. The Pawsey Supercomputing Research Centre's Quantum Supercomputing Innovation Hub and this work was made possible by a grant from the Australian Government through the National Collaborative Research Infrastructure Strategy (NCRIS). The authors acknowledge the collaboration with QuEra within the Quantum Alliance program, which provided technical support and access to neutral-atom quantum computing resources.

\bibliographystyle{IEEEtran}
\bibliography{sn-bibliography}% common bib file

\newpage

\appendix

\section{Rydberg Hamiltonian Formulation}\label{app:hamiltonian}

For completeness, we briefly present the quantum mechanical formulation underlying neutral-atom processors. While the main text focuses on the high-level mapping of STIN optimization problems to graph structures, the following describes how these problems are represented at the hardware level.

Neutral-atom devices encode qubits in single atoms trapped by optical tweezers. Information is stored in their electronic states, and interactions arise when atoms are driven into highly excited \emph{Rydberg states} $\ket{1}$. Neighboring atoms in these states experience strong van der Waals interactions, leading to the so-called \emph{Rydberg blockade}: two adjacent atoms cannot be simultaneously excited. This physical mechanism naturally enforces independent set constraints.

The dynamics of the system are described by the Rydberg Hamiltonian:
\begin{equation}\label{eq:hamiltonian}
    \frac{\mathcal{H}}{\hbar} = \frac{\Omega(t)}{2}\sum_{i=1}^{N} \left( |0_i\rangle \langle 1_i| + |1_i\rangle \langle 0_i| \right)
    - \sum_{i=1}^{N} \Delta_i(t) \hat{n}_i
    + \sum_{j<k} \frac{C_6}{|\vec{r}_j - \vec{r}_k|^6} \hat{n}_j \hat{n}_k,
\end{equation}
where:
\begin{itemize}
    \item $\Omega(t)$ is the Rabi frequency controlling the laser-driven coupling between atomic states,
    \item $\Delta_i(t)$ is the atom-specific detuning,
    \item $\hat{n}_i = |1_i\rangle \langle 1_i|$ is the Rydberg occupation operator,
    \item and the last term encodes the interaction between atoms at positions $\vec{r}_j$ and $\vec{r}_k$ with van der Waals coefficient $C_6$.
\end{itemize}

By arranging atoms in a two-dimensional geometry, the interaction radius can be tuned to reproduce the adjacency structure of a target graph. As a result, the low-energy states of $\mathcal{H}$ correspond to independent sets of that graph, and with appropriate detuning and weighting, to Maximum Weight Independent Set (MWIS) solutions. This property forms the basis for mapping the STIN optimization problems onto neutral-atom quantum hardware.

\section{Embedding Optimization Problems on the QPU}\label{app:embedding}

The first step in applying quantum optimization algorithms on the target \textit{Aquila} platform~\cite{wurtz2023aquila} is to embed the target combinatorial problem into the physical qubit register. Formally, we consider an input graph instance $\mathcal{G}(\mathcal{V}, \mathcal{E})$, where $\mathcal{V}$ is the set of vertices and $\mathcal{E}$ the set of edges. The goal is to map $\mathcal{G}$ into a UDG~\cite{clark1990unit}, where each vertex $i \in \mathcal{V}$ is assigned to a position $\vec{r}_i \in \mathbb{R}^2$ representing an atom (qubit) in the register. We denote by $\mathcal{P} = \{\{i,j\} \mid i,j \in \mathcal{V}, i \neq j\}$ the set of all vertex pairs.

This embedding must respect the hardware constraints imposed by \textit{Aquila}’s physical qubit architecture. In particular, feasible embeddings must satisfy row-spacing, atom-spacing, and register-size requirements.  

To address this embedding challenge, we employ the DEN model~\cite{vercellino2023neural, vercellino2024harnessing}, specifically adapted to \textit{Aquila}’s rectangular register geometry. The DEN learns a transformation that maps an initial (possibly infeasible) set of coordinates $\{\vec{r}_i^{\,0}\}_{i \in \mathcal{V}}$ into feasible coordinates $\{\vec{r}_i\}_{i \in \mathcal{V}}$, compliant with adjacency, non-adjacency, and row-spacing constraints.

The model leverages neural networks’ ability to approximate nonlinear functions, using a modified autoencoder where the hidden \emph{coordinates layer} encodes the feasible positions $\vec{r}_i$. The network outputs squared pairwise and row distances, and optimizes them through a custom loss function, the ELF, which enforces the UDG constraints while maximizing the \emph{adjacency gap}—the difference between the closest non-adjacent pair $(i,j) \notin \mathcal{E}$ and the furthest adjacent pair $(i,j) \in \mathcal{E}$.

\medskip
\noindent\textbf{Position initializations.} Since the proposed optimization framework relies on learning spatial transformations, initial coordinates $\vec{r}_i^{\,0}$, $\forall i \in \mathcal{V}$, must be provided. A preprocessing phase therefore computes these initial positions. The purpose of this step is to support the convergence of the optimization algorithm by generating $\vec{r}_i^{\,0}$, $\forall i \in \mathcal{V}$, that approximately satisfy some of the hardware constraints. In this way, the DEN model converges in fewer iterations (epochs) than it would with purely random initialization.

For this purpose, we adopt the \textit{Fruchterman–Reingold} force-directed layout algorithm~\cite{fruchterman1991graph}. This method requires no initial coordinates and models both attractive and repulsive forces between vertex pairs according to the adjacency structure. In this formulation, repulsive forces act on all vertex pairs with magnitude
\[
\frac{k^2}{\| \vec{r}_i^{\,0} - \vec{r}_j^{\,0}\|_2^2},
\]
while attractive forces act only on adjacent pairs with magnitude
\[
\frac{\| \vec{r}_i^{\,0} - \vec{r}_j^{\,0}\|_2}{k}.
\]
The parameter $k$ determines the equilibrium distance at which the two forces balance for adjacent vertices~\cite{hagberg2008exploring}. In our setting, we choose $k = 7 \,\mu\text{m}$, which lies in the range $[D_{\min}, D_{\text{adj}}]$.
From the hardware constraints, we impose a minimum distance $D_{\min} = 4 \,\mu\text{m}$, while we set $D_{\text{adj}} = 10 \,\mu\text{m}$ as a realistic maximum separation between adjacent vertices. Larger distances would require longer QPU coherence times to reach the Rydberg blockade effect, whereas the current machine coherence time is limited to $\approx 3 \,\mu\text{s}$.

Although the Fruchterman–Reingold method does not guarantee that all constraints are satisfied in the initialization $\vec{r}_i^{\,0}, \forall i \in \mathcal{V}$, it provides a practical strategy for generating reasonable embeddings in the absence of prior coordinate information—an assumption that holds for most graphs in UDG-related applications. The algorithm is iterative and converges quickly in practice; we limit its execution to at most $1000$ iterations.

\medskip
\noindent\textbf{Distance Encoder Network method.} After the preprocessing phase, the core of the optimization framework takes place: the initialization and training of the DEN model. While the training algorithm follows the standard neural network procedure (forward pass to compute outputs and gradients, backward pass to update weights), its interpretation is specific to this context. The DEN model learns a spatial transformation that maps an initially infeasible unit-disk representation into a feasible one, while also maximizing the adjacency gap.

Importantly, training is performed independently for each graph instance. The network architecture is therefore tailored to the problem dimensionality $N = |\mathcal{V}|$ and to the adjacency structure determined by $\mathcal{E}$. For each instance, the maximum number of training epochs is fixed to $5000$.

The DEN model also includes dropout layers for regularization~\cite{wager2013dropout}. Thus, each epoch consists of two steps:
\begin{itemize}
    \item \textbf{Training step:} dropout is applied by randomly disabling neurons in hidden layers; the ELF is computed; weights are updated using the AdamW optimizer~\cite{loshchilov2017decoupled} with learning rate $lr=1\cdot 10^{-2}$.
    \item \textbf{Inference step:} dropout is disabled, and the updated embedding configuration is computed without further weight updates.
\end{itemize}

The DEN architecture consists of two main components: a \textit{trainable autoencoder} and a \textit{distance computation block}.

\medskip
\noindent\textbf{Trainable autoencoder.}  
This component encodes the initial coordinates $\vec{r}_i^{\,0}, \forall i \in \mathcal{V}$ and outputs transformed coordinates $\vec{r}_i, \forall i \in \mathcal{V}$. Coordinates are flattened into one-dimensional vectors: the input vector $I$ and the output vector $O$,
\[
I(k) = 
\begin{cases}
	\vec{r}_{k}^{\,0(x)} & 0 \leq k < N,\\[4pt]
    \vec{r}_{k-N}^{\,0(y)} & N \leq k < 2N,
\end{cases}
\qquad
O(k) = 
\begin{cases}
	\vec{r}_{k}^{(x)} & 0 \leq k < N,\\[4pt]
    \vec{r}_{k-N}^{(y)} & N \leq k < 2N.
\end{cases}
\]

All fully connected layers include bias terms and are equipped with dropout (with probability $p_{\text{drop}} = 0.3$). The last fully connected $CoordL$ layer uses a bounded activation to ensure that coordinates lie in the rectangular domain $L_x \times L_y$. Table~\ref{tab:autoencoder} summarizes the autoencoder architecture.

\begin{table*}[htbp]
\caption{Trainable autoencoder architecture for a graph with $N$ vertices.}
\label{tab:autoencoder}
\centering
\begin{tabular}{ccccc}
\toprule
& Layer type & Input size & Output size & Activation \\
\midrule
Encoder & Fully connected & $2N$ & 64 & ReLU \\
        & Fully connected & 64 & 36 & ReLU \\
        & Fully connected & 36 & 18 & ReLU \\
        & Fully connected & 18 & 9 & ReLU \\
\midrule
Decoder & Fully connected & 9 & 18 & ReLU \\
        & Fully connected & 18 & 36 & ReLU \\
        & Fully connected & 36 & 64 & ReLU \\
        & Fully connected & 64 & $2N$ & $\frac{L}{2}\tanh$ \\
\bottomrule
\end{tabular}
\end{table*}

\medskip
\noindent\textbf{Distance computation block.}  
The second component computes squared pairwise distances
\[
d_{ij}^2 = \| \vec{r}_i - \vec{r}_j \|_2^2, \quad \forall \{i,j\} \in \mathcal{P},
\]
and squared row distances
\[
(y_i - y_j)^2, \quad \forall \{i,j\} \in \mathcal{P}.
\]
This ensures that the autoencoder’s outputs are interpreted as Cartesian coordinates and provides the proper input to the loss function.

The computation is realized through two fixed-weight layers:
\begin{itemize}
    \item the \textit{difference layer} ($DiffL$), with weights in $\{\pm 1,0\}$, computes all pairwise coordinate differences;
    \item the \textit{distance layer} ($DistL$), with weights in $\{0,1\}$, combines squared differences into squared Euclidean and row distances.
\end{itemize}

Formally, $DiffL$ outputs are
\begin{align}
u_{i(N-1)-\binom{i}{2}+j-i-1} &= \vec{r}_{i}^{\,(x)} - \vec{r}_{j}^{\,(x)} & \forall \{i,j\} \in \mathcal{P}, \\
u_{(N-1)(\frac{N}{2} + i)-\binom{i}{2}+j-i-1} &= \vec{r}_{i}^{\,(y)} - \vec{r}_{j}^{\,(y)} & \forall \{i,j\} \in \mathcal{P}.
\end{align}
After squaring, $DistL$ outputs
\[
v_k = 
\begin{cases}
u^{2}_k + u^{2}_{\binom{N}{2}+k} & 0 \leq k < \binom{N}{2}, \\
u^{2}_k & \binom{N}{2} \leq k < 2\binom{N}{2}.
\end{cases}
\]

\begin{figure}[ht!]
\centering
\includegraphics[width=\textwidth]{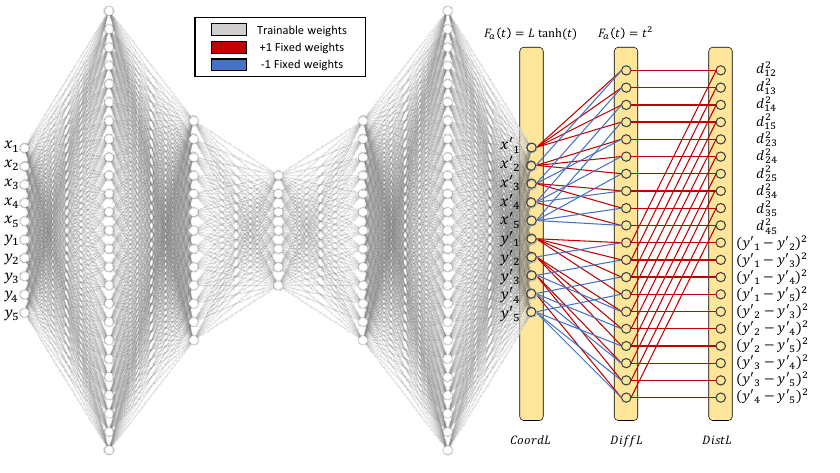}
\caption{DEN architecture for a 5-vertex embedding example.}
\label{fig:DEN}
\end{figure}

To guide the DEN model towards feasible embeddings on the \textit{Aquila} QPU, we define a custom loss function, hereafter referred to as the ELF. This loss enforces the hardware constraints on distances while promoting UDG feasibility and maximizing the adjacency gap.

The function takes as input the squared pairwise distances and row distances produced by the DEN, together with the adjacency tensor of the target graph, determined by $\mathcal{E}$. For each $\{i,j\} \in \mathcal{P}$, the ELF considers:

\begin{itemize}
    \item \textbf{Minimum distance constraint:}  
    Adjacent vertices must be at least $D_{\min}=4\,\mu\text{m}$ apart, while non-adjacent pairs should be separated by more than the maximum feasible adjacency distance $D_{\text{adj}} = 10\,\mu\text{m}$. A margin-based ranking loss penalizes violations:
    \[
    \mathcal{L}_{\min} = \sum_{\{i,j\}\in\mathcal{P}}
    \max\big(0,\, d_{\text{target},ij}^2 - d_{ij}^2\big),
    \]
    where $d_{ij}^2$ is the squared distance predicted by the network, and $d_{\text{target},ij}$ equals $D_{\min}$ (if $\{i,j\}\in \mathcal{E}$) or $D_{\text{adj}}$ (if $\{i,j\}\notin \mathcal{E}$).
    
    \item \textbf{Maximum distance constraint:}  
    To control the maximum separation of adjacent vertices, a complementary ranking loss is introduced:
    \[
    \mathcal{L}_{\max} = \sum_{\{i,j\}\in\mathcal{P}}
    \max\big(0,\, d_{ij}^2 - d_{\text{target},ij}^2\big),
    \]
    where $d_{\text{target},ij}$ is set to $D_{\text{adj}}$ for adjacent pairs, and to an upper bound on the register diagonal, defined as $\max(L_x,L_y)\cdot \sqrt{2}$, for non-adjacent ones.
    
    \item \textbf{Row spacing constraint:}  
    The vertical separation between rows must be at least $D_{\text{row}} = 2\,\mu m$. This is enforced by a polynomial penalty on squared row distances:
    \[
    \mathcal{L}_{\text{row}} = \sum_{\{i,j\}\in\mathcal{P}}
    \max\!\left(0,\; -\tfrac{4}{D_{\text{row}}^2}(y_i-y_j)^4 + 4(y_i-y_j)^2\right).
    \]
    
    \item \textbf{Unit-disk margin (adjacency gap):}  
    Finally, the ELF promotes separability between adjacent and non-adjacent pairs by maximizing the distance gap:
    \[
    \mathcal{L}_{\text{ud}} = \max_{\{i,j\}\in \mathcal{E}} d_{ij}^2 \;-\; 
    \min_{\{i,j\}\notin \mathcal{E}} d_{ij}^2.
    \]
\end{itemize}

The overall loss is defined as a weighted sum of these contributions:
\[
\mathcal{L} = \mathcal{L}_{\min} + \mathcal{L}_{\text{row}} +
\begin{cases}
    \mathcal{L}_{\max} + \mathcal{L}_{\text{ud}}, & N \leq 20,\\[6pt]
    0.1 \,\mathcal{L}_{\text{ud}}, & N > 20.
\end{cases}
\]

This formulation balances hard feasibility constraints (minimum spacing, row separation) with soft optimization objectives (adjacency gap maximization). For small graphs ($N \leq 20$), all terms are fully enforced; for larger graphs, a relaxed weighting of the adjacency gap is used to support convergence within the fixed epoch budget.

\medskip
\noindent\textbf{Refined embedding adjustment.} Once an initial embedding is obtained from the DEN model, the vertex positions are further refined through a continuous constrained optimization procedure. First, all coordinates are translated so that the lower-left corner of the layout aligns with $(0,0)$, and positions are rounded to a fixed grid to simplify constraint enforcement.

Safe bounding boxes are then computed for each vertex, enforcing the minimum Euclidean distance constraint ($d_{\min}=4\,\mu m$), row spacing ($d_{\text{row}}=2\,\mu m$), and register area limits ($76\,\mu m \times 128\,\mu m$). These bounding boxes define the feasible search space for vertex positions, ensuring that hardware and geometric constraints are respected during optimization.

Within this feasible region, a continuous optimization problem is solved using the L-BFGS-B algorithm~\cite{zhu1997algorithm}. The objective function is a margin-based loss that penalizes adjacent vertices placed too far apart and non-adjacent vertices placed too close, effectively improving adjacency separation while maintaining safe distances for non-edges. The optimization returns a refined set of coordinates, which are then rounded to a fixed precision, compatible with \textit{Aquila}'s QPU requirements, and validated.

Validation is performed by comparing the largest distance among adjacent vertices $d$, with the smallest distance among non-adjacent vertices $D$. This defines a feasible interval $[d,D]$ for a unit-disk realization: if $d<D$, the graph admits a unit-disk embedding; otherwise, it is marked as non-unit-disk. This procedure ensures that adjacency constraints, row spacing, and register boundaries are all satisfied, producing a high-quality embedding suitable for deployment on the Aquila QPU and for subsequent analysis or visualization.

\section{Quantum Optimization with Aquila}

After computing refined embeddings for the STIN optimization graphs, the vertex positions are directly mapped to qubit locations on the \textit{Aquila} QPU. Each vertex corresponds to a single neutral atom, whose $x$-$y$ position in the tweezer array is given by the embedding $\{\vec{r_i}\}_{i\in \mathcal{V}}$. This mapping provides the geometric basis for encoding adjacency constraints through the Rydberg blockade, enabling a direct realization of graph-based combinatorial optimization problems.

\medskip
\noindent\textbf{Quantum Adiabatic Algorithm for MWIS problems.} The MWIS problem, central to the SSP, is solved on \textit{Aquila} using the QAA~\cite{albash2018adiabatic}. Vertex weights are represented through locally addressable detuning of each qubit ($\Delta_i$ in Eq.\eqref{eq:hamiltonian}), while adjacency constraints are enforced by the Rydberg blockade between nearby atoms.

The mapping from embedding geometry to quantum control parameters relies on the \emph{Rydberg blockade radius} $r_b$, which sets the effective interaction range. It is given by:
\begin{equation}
    r_b = \sqrt[6]{ \frac{C_6}{(2\Omega)^2 + \Delta^2}},
\end{equation}
where $C_6$ is the van der Waals coefficient, $\Omega$ the global Rabi frequency, and $\Delta$ the global detuning~\cite{pichler2018quantum}. Following the guideline of~\cite{bombieri2025quantum}, we impose the operational condition
\[
\frac{\Delta}{3} \approx \Omega,
\]
which balances interaction strength and adiabatic passage speed.

Since embeddings provide a feasible distance interval $[d, D]$ between vertices, we select a working blockade radius as the geometric mean~\cite{matwiejew2025continuous}
\[
r_b = \sqrt{d \cdot D}.
\]
This avoids pathological cases where $r_b$ would otherwise lie too close to either $d$ or $D$, ensuring a valid physical encoding when the graph is a proper UD instance, while maintaining feasible control parameters for non-UD embeddings.

The \textit{Aquila} hardware provides two detuning channels: a uniform \emph{global detuning} $\Delta_{\text{global}}$ and a spatially resolved \emph{local detuning} $\Delta_{\text{local}}$. For each qubit $i$, the effective detuning is
\begin{equation}
    \Delta_i = \Delta_{\text{global}} + \Delta_{\text{local},i}.
\end{equation}

The global channel sets the baseline sweep, while the local channel encodes vertex weights into site-dependent energy shifts. Because the \textit{Aquila} platform only supports non-positive values for the local channel ($\Delta_{\text{local},i} \leq 0$), the weights are rescaled such that the vertex with the lowest weight receives the most negative detuning, while higher weights correspond to values closer to zero. In this way, heavier vertices become energetically favored during the adiabatic evolution, correctly biasing the dynamics toward high-weight independent sets.

\begin{figure}[H]
    \centering
    \includegraphics[width=1\linewidth]{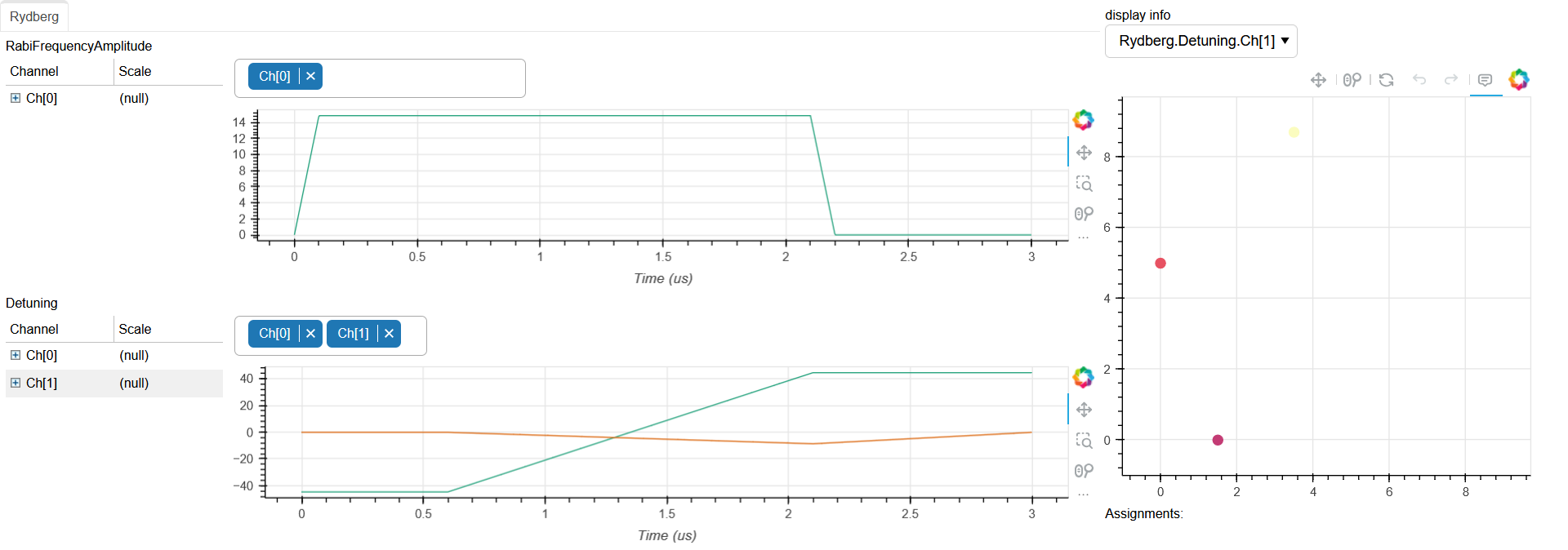}
    \caption{Adiabatic pulse schedule on Aquila for a three-vertex graph. Global Rabi and detuning sweeps are overlaid with local detuning offsets (color-coded by vertex weight).}
    \label{fig:pulses}
\end{figure}

Figure~\ref{fig:pulses} shows an example pulse sequence: the global Rabi frequency $\Omega(t)$ and detuning $\Delta_{\text{global}}(t)$ are varied piecewise-linearly in time, while the local channel $\Delta_{\text{local},i}$ introduces vertex-specific shifts. The combination encodes the weighted graph structure into the adiabatic schedule executed on \textit{Aquila}.

For small graphs ($\leq 15$ vertices), the QAA is simulated locally on the emulator from the \texttt{bloqade} SDK\footnote{\url{https://bloqade.quera.com/latest/analog/}}. Larger graphs are submitted to \textit{Aquila}’s cloud interface, where experimental counts are collected and post-processed to identify maximum-weight independent sets.

The total duration of our QAA protocol is $3~\mu\text{s}$. The annealing schedules are implemented as piecewise-linear functions, with segments
\[
\text{dur}_1 = [0.1, 2.0, 0.1, 0.8]~\mu\text{s}\quad\text{and}\quad
\text{dur}_2 = [0.6, 1.5, 0.9]~\mu\text{s},
\]
as shown in Fig.~\ref{fig:pulses}. These durations were chosen to create reasonably long ramps in the control pulses, in compliance with the principle of adiabatic computing, ensuring the system evolves slowly enough to remain close to the instantaneous ground state throughout the anneal.

\medskip
\noindent\textbf{Postprocessing and solution refinement.} The output of the QAA is inherently probabilistic: each run returns a bitstring sampled from the final quantum state. To obtain meaningful statistics, we perform $300$ shots per graph instance, generating a distribution over candidate solutions. Several sources of imperfection affect this distribution. On the modeling side, embeddings that are not strictly unit-disk may lead to mismatches between the intended graph $\mathcal{G}(\mathcal{V}, \mathcal{E})$ and its hardware realization, causing missing or spurious edges $\mathcal{E}'$. On the device side, quantum noise—including measurement errors and stray external fields—can bias the observed bitstring frequencies away from the ideal distribution.

Postprocessing addresses these issues by extracting feasible and competitive solutions from the noisy samples. The first step is to interpret \textit{Aquila}’s measurement convention: atom traps detected as empty correspond to the Rydberg state $\ket{1}$, which is denoted by logical ``0,'' and in our mapping this indicates that the associated vertex is \emph{selected} in the independent set. Conversely, a trapped atom (logical ``1'') means the vertex is excluded. Each bitstring is thus converted into a candidate independent set.

From these candidates, a refinement procedure is applied. Bitstrings that violate independence are discarded. The remaining sets are then improved with a greedy augmentation step: unselected vertices are inspected in descending order of weight, and each is added whenever it does not break independence. This ensures that feasible solutions are promoted toward maximum-weight independent sets, compensating for both quantum noise and imperfect embeddings.

\begin{figure}[H]
    \centering
    \includegraphics[width=1\linewidth]{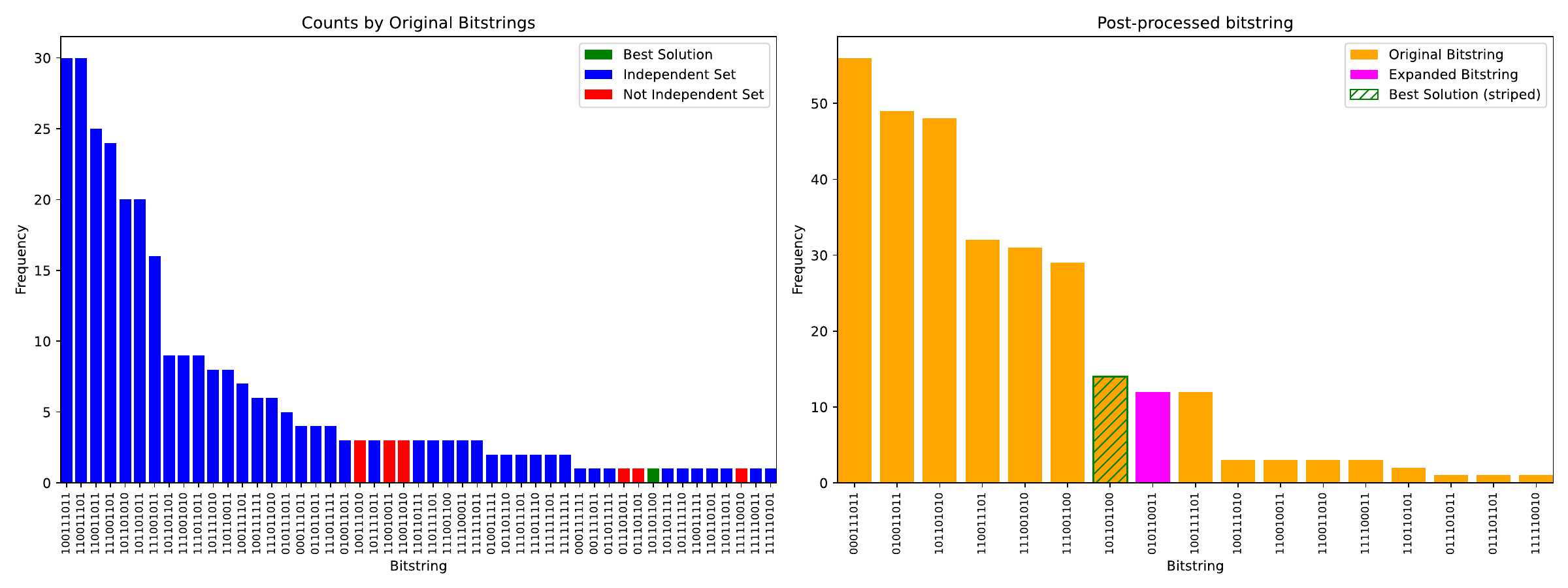}
    \caption{Histograms of QAA output before (left) and after (right) postprocessing. In the raw distribution, bitstrings are classified as independent, non-independent, or best solution. After refinement, bitstrings are distinguished according to whether they were directly sampled or obtained via greedy augmentation.}
    \label{fig:histograms}
\end{figure}

Figure~\ref{fig:histograms} illustrates this process for a representative instance. The left panel shows the raw QAA distribution, with bitstrings colored according to whether they represent independent sets, non-independent sets, or the best observed solution. The right panel shows the postprocessed distribution, where refined bitstrings are highlighted differently depending on whether they were already present in the raw samples or obtained through classical refinement. As can be observed, the refinement increases the occurrence of the best solution by an order of magnitude.

By combining probabilistic quantum sampling with deterministic classical correction, this hybrid pipeline ensures that the final solutions are both feasible and competitive, enabling their use as reliable inputs for subsequent STIN optimization tasks.

\end{document}